\documentclass[aps,pre,reprint,floatfix,longbibliography,superscriptaddress]{revtex4-1}
\pdfoutput=1 \usepackage{graphicx} \usepackage{amssymb,amsfonts,amsmath}
\usepackage{braket} \usepackage{color} \usepackage{multirow}

\newcommand{\e}{\varepsilon}
\newcommand{\dd}{\partial}
\newcommand{\Gd}{\Psi^\dagger}
\newcommand{\dotprod}[2]{\braket{ #1 | #2}}
\newcommand{\chern}{{\cal C}}
\newcommand{\kv}{{\mathbf{k}}}

\begin{document} \title{Topological protection can arise from thermal fluctuations and interactions}

\author{Ricardo Pablo Pedro}
\thanks{These authors contributed equally to this work.}
\affiliation{Department of Chemistry, Massachusetts Institute of Technology, Cambridge, MA 02139, USA}

\author{Jayson Paulose}
\thanks{These authors contributed equally to this work.}
\affiliation{Department of Physics and Institute of Theoretical Science, University
  of Oregon, Eugene, OR 97403, USA}
  
\author{Anton Souslov}
\affiliation{The James Franck Institute and Department of Physics, 
The University of Chicago, Chicago, IL 60637, USA}
\affiliation{Department of Physics, University of Bath, Bath BA2 7AY, United Kingdom}

\author{Mildred Dresselhaus}
\affiliation{Department of Physics and Department of Electrical Engineering and Computer Science, Massachusetts Institute of Technology, Cambridge, MA 02139, USA}

\author{Vincenzo Vitelli}
\email{vitelli@uchicago.edu}
\affiliation{The James Franck Institute and Department of Physics, 
The University of Chicago, Chicago, IL 60637, USA}

\begin{abstract}
Topological quantum and classical materials can exhibit robust properties that are protected
  against disorder, for example for noninteracting particles and linear waves.
  Here, we demonstrate
  how to construct topologically protected states that arise from the
  combination of strong interactions and thermal fluctuations inherent to soft
  materials or miniaturized mechanical structures. Specifically, we consider fluctuating lines under tension (e.g., polymer or vortex lines), subject to a class of spatially modulated substrate potentials. At equilibrium, the lines acquire
  a collective tilt proportional to an integer topological invariant called the
  Chern number. This quantized tilt is robust against substrate disorder,
  as verified by classical Langevin dynamics simulations. 
  This robustness arises because excitations in this system of thermally fluctuating lines 
  are gapped by virtue of inter-line interactions. We establish the
  topological underpinning of this pattern via a mapping that we develop 
  between the interacting-lines system and a hitherto unexplored generalization
  of Thouless pumping to imaginary time.  Our work points to a new class of classical topological phenomena
  in which the topological signature manifests itself in a structural property observed at finite temperature
  rather than a transport measurement. 
\end{abstract}

\maketitle

Topological mechanics~\cite{Huber2016, Susstrunk2015, Kane2014,Paulose2015,Prodan2009,Nash2015} and 
 optics~\cite{Rechtsman2013,Lu2014} typically focus on systems of linear waves assuming that mode
 interactions and finite-temperature effects can be ignored in deriving the relevant topological invariants and corresponding physical observables.
 However, these assumptions break-down when structures are miniaturized down to the micron scale. The resulting interplay between large-amplitude thermal displacements and mechanical constraints arises in contexts ranging from molecular robotics to soft materials. In this Letter, we show that thermal fluctuations and interactions, far from being a hindrance,
can actually create topologically protected states by acting in tandem. We provide a specific illustration of this mechanism in thermally fluctuating and interacting lines (or chains) under tension whose statistics describe such diverse systems as directed polymers~\cite{DeGennes1968,LeDoussal1991,Nelson2002,Rocklin2012}, crystal step edges on vicinal surfaces~\cite{Bartelt1990} and vortex lines in superconductors~\cite{Nelson1993,Polkovnikov2005}. 

Consider, as an example, polymers confined
within a thin layer parallel to the $xy$--plane and experiencing a tension $\tau$ along the direction $y$
(Fig.~1a). Spatial modulations in the polymer-substrate interaction potential influence the
equilibrium monomer density along the chain (Fig.~1b). Previous studies of
directed polymer systems have focused on the effect of localized or
randomly-distributed constraining potentials on line conformations~\cite{Nelson1993,Dotsenko2010}. By contrast,
we characterize the patterns of directed lines induced by \emph{periodic} substrate
potentials. Although the underlying principle is more general, we focus here on
the specific form for the potential energy per unit
length~\cite{Wang2013a,Wei2015}
\begin{equation} \label{eq:weimueller} V(x,y) = V_1 \cos \left(\frac{2\pi q}{a}x
\right) + V_2 \cos \left(\frac{2\pi p}{a}x - \frac{2\pi}{\lambda}y\right),
\end{equation}
which combines a $y$-independent sinusoidal component (first term) with a mixed one
(second term) which slides along the $x$-direction as $y$ advances (Fig.~1c). The
period in the $x$--direction is given by $a$ divided by the greatest common
divisor of the integers $p$ and $q$, and in the $y$--direction the period is
denoted by $\lambda$. ($V_1$ and $V_2$ are substrate energy scales.) The
form of the potential in Eq.~(\ref{eq:weimueller}) is motivated by an
analogy between the system of fluctuating lines and the so-called Thouless charge
pump~\cite{Thouless1983}, which was recently realized and extended in ultracold
atom experiments~\cite{Nakajima2015,Lohse2015,Lohse2018}.
As we shall see, the formalism of the Thouless pump
needs to be extended to account for the thermally fluctuating classical
systems considered here.

\begin{figure}[t!]
  \includegraphics{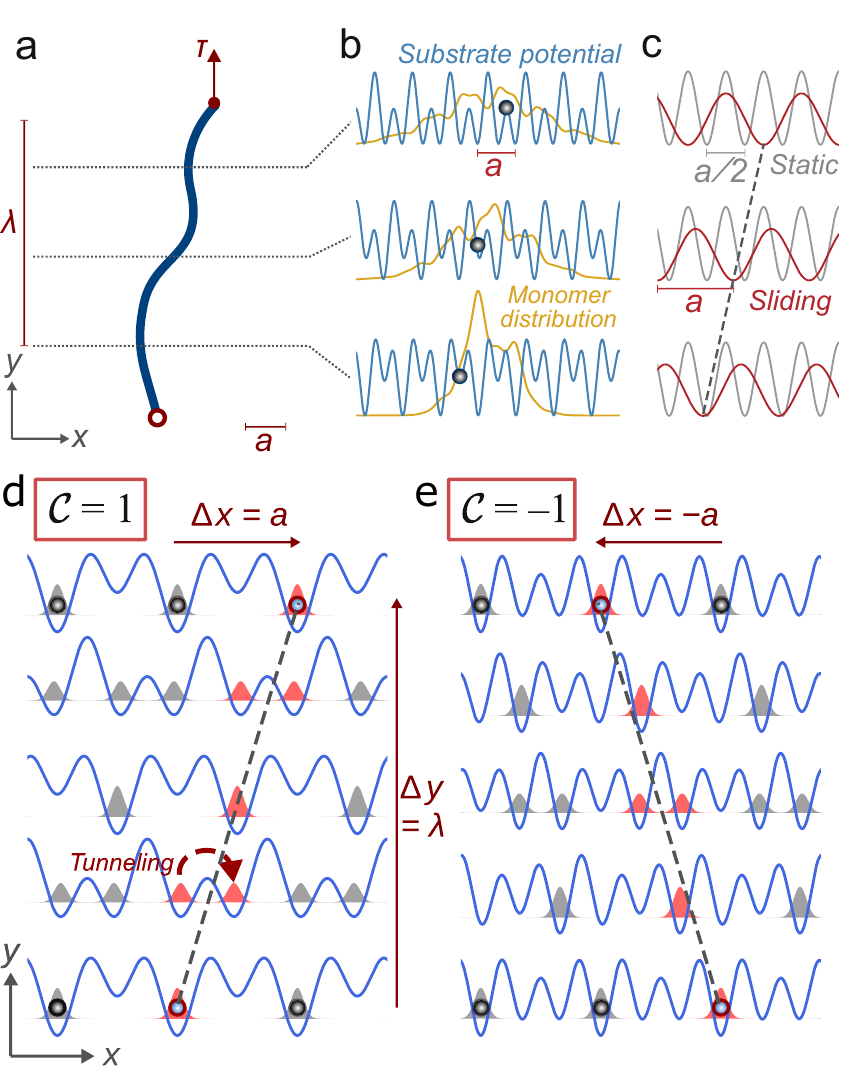}
  \caption{{\bfseries Directed lines and doubly-periodic substrate
      potentials.} (a) Schematic of single directed line in a potential described by Eq.~(1)
    with $(p,q)=(1,2)$, $y$-axis period $\lambda$, and $x$-axis period $a$. 
   (b) Substrate potential (blue curves) and the theoretical density distribution 
    [from Eq.~(\ref{eq:schrod}), 
    yellow curves] for a single chain at three $y$-positions indicated
    by the dotted lines. 
   (c) The compound potential $V(x,y)$ [blue]
    combines two components, static (gray) with wavelength
    $a/q=a/2$ and sliding (red) with wavelength $a/p=a$.
    (d) Illustration of a
    Thouless pump for a potential with $(p,q)=(1,2)$, corresponding to
    $\mathcal{C}=1$. Under a filling density of one electron per lattice
    constant, each electron is exponentially localized to a unique unit cell.
    The drift of one such localized wavefunction over an adiabatic cycle is
    shown schematically; it is exactly quantized to $\mathcal{C}$ steps of
    lattice size $a$ over each period $\lambda$ of the potential variation along
    the $y$ direction. The tunneling of probability weight between adjacent
    potential minima during the adiabatic evolution, indicated by the dashed
    arrow, is crucial for the shift.
    (e) Same as (d) for a potential with $(p,q)=(2,3)$ for
    which  $\mathcal{C}=-1$.}
  \label{fig:mott}
\end{figure}

A quantum Thouless pump describes the adiabatic flow of charge in a
one-dimensional electron gas subject to a potential that varies periodically in
both space and time. When the electrons populate an energy band completely, the
number of electrons transported in one cycle is quantized to an integer-valued
topological invariant of the filled band---the Chern number, $\chern$~\cite{Thouless1982}. The
static potential in Eq.~\ref{eq:weimueller} can be viewed as a time-dependent
potential with the spatial coordinate $y$ interpreted as the time coordinate. For electrons
experiencing this potential, the Chern numbers are determined by the integers
$p$ and $q$~\cite{Avron2014} and can be nonzero, leading to charge flow. For the
potential in Fig.~1 with $(p,q) = (1,2)$, the lowest band has $\chern = 1$.
Hence, under a filling density of one electron per lattice period $a$,
the electrons are shifted to the \emph{right} by one lattice period over one
time cycle $\lambda$, see Fig.~1d. By contrast, Fig.~1e illustrates the case
$(p,q) = (2,3)$, for which $\chern = -1$. In this case of so-called
``anomalous'' pumping~\cite{Wei2015}, the electrons flow to
the \emph{left} even though the potential is still sliding to the right. As long
as the gap between occupied and unoccupied bands remains open, the topological
nature of $\chern$ insures that the charge flow is robust against electron
interactions and disorder in the potential $V(x,y)$~\cite{Niu1984}.

Can we formulate a thermal generalization of Thouless pumping and use it to
engineer topological soft materials? 
Here, we show that directed fluctuating lines can order into tilted patterns that mimic the
spacetime paths traced by the quantum particles in Fig.~1d--e. Several studies have
shown that the conformations of a thermally fluctuating chain can be mapped to
the paths of a quantum
particle~\cite{DeGennes1968,LeDoussal1991,Kamien1992,Nelson2002,Rocklin2012,Rocklin2013}.
However, Thouless pumping introduces a new facet to this mapping: the
requirement of a gapped phase. For electrons, the gapped phase is accomplished
by filling a band, which requires Pauli exclusion---a distinctive feature of
fermions. To recreate exclusion effects in classical fluctuating lines, we exploit the
fact that the chains do not cross. Remarkably, the noncrossing constraint
reproduces the effects of Fermi statistics in the directed line 
system~\cite{DeGennes1968}, allowing us to ``fill'' bands by tuning the number
of chains per lattice constant along the $x$-direction.

\begin{figure*}[t!]
  \includegraphics{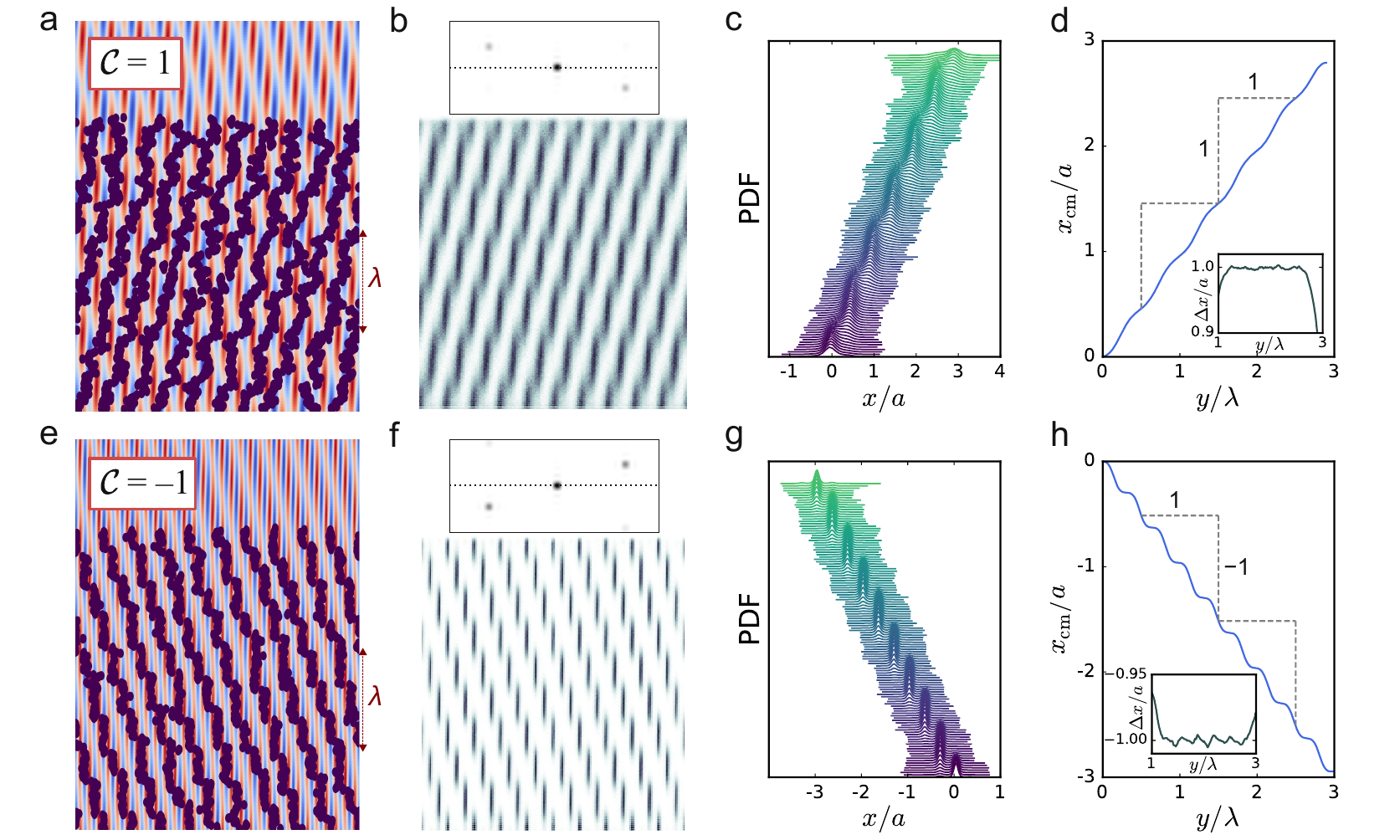} 
  \caption{{\bfseries Topological tilt of directed line conformations.} {\bfseries a,}
    Snapshot of a molecular dynamics simulation (see SI) of ten noncrossing
    directed lines experiencing the substrate potential from
    Fig.~2a with $\mathcal{C}=1$ under commensurate filling (one
    chain per unit cell of the potential along the $x$ direction).
    {\bfseries b,} Equilibrium monomer density distribution, and
    numerically-computed scattering intensity profile (inset; see Methods). 
    {\bfseries c,} Probability
    distribution of monomer $x$-position for a subset of monomers along the
    length of the chains. Data from different chains are aggregated by first
    shifting the $p$th chain by an amount $pa$ along $x$, where $p \in \{0,...,9\}$
    indexes the chains in order from left to right. {\bfseries d,} Centre of
    mass computed from the probability distributions in {\bfseries c}. The inset
    shows the shift over one period, $x_\text{cm}(y)-x_\text{cm}(y-\lambda)$, which agrees
    with the prediction of $\mathcal{C}=1$ away from the line ends.
    {\bfseries e--h,} Same as {\bfseries a--d} for the potential from
    Fig.~2b with $\mathcal{C}=-1$. The lines display an anomalous tilt to the left,
    even though the potential slides to the right with increasing $y$.
  } \label{fig:pumping}
\end{figure*}

To test whether interacting, thermal chains can replicate topological
charge pumping, we have conducted Langevin dynamics simulations of collections
of chains interacting with each other via a harmonic contact repulsion
below a cutoff separation
and, in addition, interacting with the substrate according to
Eq.~(\ref{eq:weimueller}) with $V_1$ and $V_2$ of the same order as the thermal
energy $k_B T$. (Simulation details are provided in SI.) We emulate filling of
the lowest band by including lines at a density of one chain per lattice
constant $a$. When parameters $(p,q) = (1,2)$ are chosen so that $\chern=1$,
the chains acquire a collective tilt to the right (Fig.~2a) which is also
apparent in the equilibrium density profile (Fig.~2b). Probability distributions
of the monomer $x$-positions at different values of $y$ show that the shift in
average chain position advances to the right by one lattice constant per cycle,
matching the quantization expected from the Chern number to within 1\% accuracy
(Figs.~2c--d).

In contrast to the quantum pump, the topological tilt of the lines is a direct
consequence of many-body interactions between the chains: a single
chain on an otherwise empty lattice diffuses freely through the system and, on
average, does not tilt, see Supplementary Movie for an illustration. Moreover,
thermal fluctuations do not destroy the topological state, but rather are
crucial for creating the tilt via a series of ``thermal tunneling'' events
visible in the density profiles of Fig.~2c. These events are analogous to the
quantum tunneling in Fig.~1.

The non-vanishing slope resulting whenever $\chern \ne 0$ cannot be intuited
from superficial aspects of the substrate potential or from the (real-time)
dynamics of classical particles under the same potential. For instance,
Fig.~2e--h shows the case $(p,q) = (2,3)$, for which $\chern = -1$.
Surprisingly, the lines tilt to the left even though the sliding part of
the potential, given by the last term in Eq.~(\ref{eq:weimueller}), has the
positive slope $a/(\lambda p)$ which by itself would suggest a tilt to the
right. Note that the topologically distinct left- and right-leaning
configurations can be differentiated by their diffraction patterns (Figs.~2b and
2f, insets), suggesting a scattering experiment that would directly measure the
underlying topological index.

To rigorously establish the topological origin of the observed tilt, we turn to
the aforementioned mathematical correspondence between quantum particles and
thermally fluctuating lines\cite{Edwards1965,DeGennes1969,Matsen2006,DeGennes1968,LeDoussal1991,Kamien1992,Nelson2002,Rocklin2012,Rocklin2013}.
This quantum-classical correspondence stems from the formal similarity between
the Schr\"odinger equation and the diffusion equation describing the chain
statistics:
\begin{equation}
  \label{eq:schrod}
  \partial_y \Psi = \frac{k_B T}{2 \tau} \partial_x^2 \Psi - \frac{1}{k_B T} V
\Psi \equiv H \Psi .
\end{equation}
Interpreted using the directed line language, Eq.~(\ref{eq:schrod}) describes the
(real) probability distribution $\Psi(x,y)$ of chain location $x$ at distance
$y$ from the constrained end at $y=0$, given the initial distribution
$\Psi(x,0)$. [The external tension
prevents directed chains from doubling back on themselves, which means
that the instantaneous chain configurations are described by single-valued
functions $x(y)$.] On the other hand, upon equating $y$ with $i t$ and $k_BT$
with $\hbar$, Eq.~(\ref{eq:schrod}) describes the evolution of the (complex)
wavefunction $\Psi(x,t)$ for a particle of mass $\tau$ in the time-dependent
potential $V(x,t)$.
The transformation to imaginary time is a key aspect of our proposal in two ways. First, it
guarantees that the solutions to Eq.~(\ref{eq:schrod}) for long chains are
described by the ground-state wavefunction of the analogous quantum system (see
SI). Below we exploit this condition, known as \emph{ground-state dominance}, to
generate a gapped state. Going to imaginary time also turns wavelike Bloch
states into states that decay over time, and thus requires an extension of
the standard formalism of Thouless pumping beyond the quantum case which we perform later on.

Inter-line interactions, together with ground-state dominance, can give rise to
gapped phases. 
To see this, consider the $y$-evolution of the joint probability distribution of
$x$-positions  $\{x_0(y),x_1(y),...,x_N(y)\}$ of $N$ chains
in a \emph{$y$-independent}  potential $V(x)$, such as the potential in
Eq.~(\ref{eq:weimueller}) when $V_2 = 0$. 
This many-body probability is described via the exchange-symmetric  eigenstates
of the effective Hamiltonian $H$ in Eq.~(\ref{eq:schrod}), augmented by a pair
interaction term of the form $(k_B T)^{-1} \sum_{i<j}V_\text{p}(x_i-x_j)$. These
(bosonic) many-body eigenstates may be challenging to describe. However, a
tremendous simplification exists for non-crossing directed lines, for which
the pair potential is infinitely large when the positions of two lines
coincide at any $y$ and is zero otherwise: $V_\text{p}=c\delta(x_i-x_j)$, $c \to
\infty$. In this case, there is a one-to-one mapping between the requisite
exchange-symmetric line eigenstates and the many-body wavefunctions of $N$
noninteracting \emph{fermions} confined to the $x$--axis and experiencing the
same substrate potential $V(x)$~\cite{Girardeau1960,DeGennes1968}. In
particular, if the number of lines is equal to the number of lattice periods,
the ground state is obtained by filling up the lowest band entirely. This
trivial electronic insulator in the fermion picture describes a Mott insulator
in the fluctuating-line picture: a state in which excitations are gapped by virtue of
interactions. The many-body joint probability distribution of the line system
$\Psi_0^\text{P}$ is then equal to the {\it absolute value} of the fermionic
ground state $\Psi_0^\text{F}$ ~\cite{Girardeau1960,DeGennes1968,Rocklin2012}.

\begin{figure}[t!]
  \includegraphics{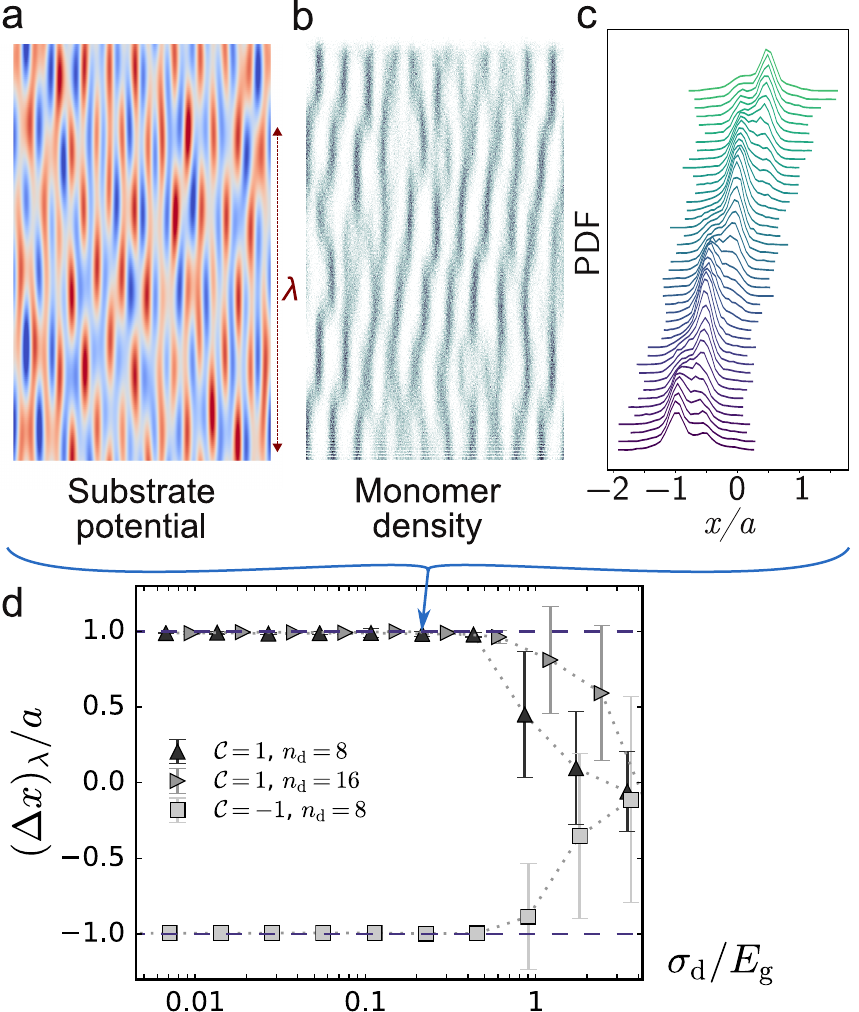}
  \caption{{\bfseries The line tilt is robust against disorder.} {\bfseries
a,} Example of a substrate potential with $\mathcal{C}=1$ from Fig.~2a, with
random disorder added. {\bfseries b,} Equilibrium monomer density distribution
for potential in {\bfseries a} under commensurate filling. {\bfseries c,}
Aggregated probability density function of monomer $x$-positions. Although the
density profiles of individual chains show deviations, the aggregated profile
maintains the quantized tilt. {\bfseries d,} Tilt as measured in simulations for
increasing disorder added to the substrate interaction for the potentials studied in
Fig.~2. Each point represents an average over ten realizations of random
disorder; the error bars represent estimated standard deviations. Triangles and
squares correspond to underlying periodic potentials with Chern numbers
$\mathcal{C} = 1$ and $-1$ respectively, with $n_\text{d}$ additional modes with
random amplitude and phase added
on. The quantized tilt is preserved until 
the disorder strength $\sigma_\text{d}$ becomes comparable to the
excitation gap $E_\text{g}$. }
  \label{fig:disorder}
\end{figure}

When this gapped state is subjected to an additional $y$-dependent potential,
such as the $V_2$ term in Eq.~(\ref{eq:weimueller}), the probability
distributions are modulated along the chain length. As long as the energy gap
remains open throughout, time-dependent perturbation theory [adapted to the
imaginary-time evolution of Eq.~(\ref{eq:schrod})] can be used to evaluate the
adiabatic change in the densities of the lines along the $y$-direction. As we show in
the SI, the instantaneous probability current across the system
can be expressed as 
$J(y) = \frac{1}{L} \partial_y \langle X \rangle$,
where crucially $\langle X \rangle$ depends only on the square modulus of the ground-state
wavefunction.
As a result, 
the density current is unchanged by the
line-fermion mapping $\Psi_0^\text{P} = |\Psi_0^\text{F}|$ and by the
transformation to imaginary time. The shift in the center of mass of the
chains over one cycle corresponds exactly to the net shift of electrons
belonging to the filled band in the Thouless pump~\cite{Thouless1983,Niu1984},
\begin{equation}
  \label{eq:chern} \frac{ \langle \Delta x \rangle_\lambda}{a} = \frac{1}{a} \int_0^\lambda
  \!J(y)\,dy =  \frac{1}{2\pi}
\int_0^\lambda \!dy\! \int_0^\frac{2\pi}{a} \!dk \,\mathcal{F}(y,k) \equiv
\mathcal{C},
\end{equation} where $\mathcal{F}(y,k) = i(\langle \partial_y
u_{k}(y)|\partial_k u_{k}(y) \rangle - \text{c.c.})$ is the Berry curvature
computed using the Bloch eigenstates $\ket{u_{k}(y)}$ of the lowest band of the 
Hamiltonian in Eq.~(\ref{eq:schrod}) with the periodic potential $V(x,y)$
evaluated at a fixed $y$ and $\mathcal{C}$ is the Chern number.

Equation~(\ref{eq:chern}) establishes the topological origin of the tilt
observed in Fig.~2. The nontrivial mapping between the directed-line and the electronic
systems is a physical consequence of two features. First, adiabatic evolution
is determined solely by changes in the instantaneous eigenstates of $H$ when the
parameter $y$ is changed, and the form of $H$ is preserved exactly on both sides
of the mapping. Second, while the Berry curvature is a property of the {\it complex}
eigenstates of the Fourier-transformed Hamiltonian, the Chern number (i.e.,
integrated Berry curvature) describes the \emph{real}-valued shift in the center
of mass of the directed-line probability distribution.
Hence, the tilt angle is a physical observable proportional to the Chern number 
which is analogous to the quantized charge transport of the electronic system. 

An important property of topological adiabatic pumps is their robustness against disorder:
since the shift in centres of mass of the single-particle states is associated
with a topological index, it is unchanged by disorder in the substrate potential
as long as the energy gap between the lowest and higher bands does not close
~\cite{Niu1984}. To test the robustness of the tilt, we add a random noise
$V_\text{d}(x,y)$ to the substrate potential (implemented as a superposition of $n_\text{d}$ sine
functions with random amplitudes and phases, see Methods). Figs. 3a--b show a
substrate potential with added disorder, and the corresponding equilibrium
monomer density. The density profile in Fig.~3b looks substantially different
from its crystalline counterpart in Fig.~2b. In the absence of disorder, the quantized collective shift
of all the chains translated to a quantized tilt in the contour of each
individual chain; this is no longer true when disorder is present. Nevertheless, the \emph{aggregated}
tilt (Fig.~3c) shows a striking regularity. The measured slope of the
equilibrium directed-line conformations over one period (Fig.~3d) remains quantized by the Chern
number until the disorder strength (the standard deviation $\sigma_\text{d}$ of
the disorder potential) becomes comparable to
the energy gap $E_\text{g}$ between the occupied 
band and the next-highest band in the energy spectrum.

The topological patterning is also robust against general interactions among
chains on top of the noncrossing constraint, which translate in the
quantum language to many-body interactions (of the same functional form) among
the fermions~\cite{Girardeau1960}. As with substrate disorder, the quantization
is unaffected as long as the excitation gap remains open when the interactions
are turned on~\cite{Niu1984}. This property is demonstrated by the results of
our simulations that employ a harmonic contact potential in addition to the noncrossing constraint (see SI).

The proposed topological phenomenon stands apart from its counterparts in optics
and mechanics in several ways. The Chern number manifests itself in a structural
property which can be measured directly from the equilibrium pattern. By
contrast, in the topological band theory of classical waves, Chern numbers only
control the number of chiral edge modes which are typically probed via the
transport of energy along the edge. Moreover, to excite an acoustic, optical or
mechanical chiral edge mode, the system must be driven at a specific frequency
corresponding to the band gap, whereas in the directed-line case there is a
notion of band filling, i.e., an effective Fermi level tuned by the chain
density.

\begin{acknowledgments} 
We thank Benny van Zuiden for programming assistance, and Vadim Cheianov,
  Michel Fruchart, Alexander Grosberg, Charles L. Kane, David R. Nelson, Philip Pincus, D. Zeb Rocklin, and Tom Witten for insightful discussions. VV was primarily supported
  by the University of Chicago Materials Research Science and Engineering
  Center, which is funded by the National Science Foundation under award number
  DMR-1420709. JP acknowledges funding from NWO through a Delta ITP
  Zwaartekracht grant. RPP gratefully acknowledges the Office of Graduate
  Education of MIT for the graduate Unitec Blue Fellowship, and the King
  Abdullah University of Science and Technology for support under contract
  (OSR-2015-CRG4-2634).
\end{acknowledgments}

\appendix
\setcounter{figure}{0}
\renewcommand{\thefigure}{A\arabic{figure}}

\section{Simulation details}

We performed 2D molecular dynamics simulations in which directed lines are
approximated as discrete chains of $N$ beads (monomers) of mass $m$ connected by nonlinear springs
with equilibrium length $l_0$ and maximum extension $l_\text{max}$, implemented
using the pair potential
\begin{equation}
  \label{eq:fene}
  V_\text{chain}(r) = -\frac{K l_\text{max}^2}{2}  \log \left[1-(r-l_0)^2/l_\text{max}^2\right]
\end{equation}
for linked beads with separation $r$. To implement the noncrossing constraint,
each bead has a finite radius $r_\text{c} = l_0$ and interacts with all other
beads via a harmonic contact potential $V_\text{c}(r) = k_\text{c}(r-l_0)^2,\ r
< l_0$. Tension is applied by pulling the topmost bead along the $y$
direction with a force $\tau$. The bottom beads are either pinned to specific
points with horizontal spacing $a$ at $y = 0$ (for simulations in Fig.~1) or
confined to $y = 0$ but free to slide along the $x$ direction (for Figs. 2
and 3). The substrate interaction $V(x,y)$ per unit length is implemented as a
position-dependent potential of strength $l_0 V(x,y)$ acting on every bead.
For all simulations, we set $m=1$, $l_0=0.3$, $K=k_\text{c}=10^3$, $\tau=10$,
$a=1$, $\lambda=25$ in simulation units. The simulation box size is $L_x=10a$
and $L_y=16\lambda$ with periodic boundary conditions along $x$. The box size
along $y$ is always many times the chain length.  

Temperature is implemented by applying a friction force with drag coefficient
$\gamma$ to all beads, and adding a random force of strength $\sqrt{2\gamma
  k_\text{B}T}$ along each dimension to each bead. In all our simulations, we
set $k_\text{B}T = 1$ and $\gamma = 0.5$ in simulation units. Newton's equations
are solved using the velocity-Verlet algorithm with time steps of length $0.001$ in
simulation units, which equates to $0.0316/\omega$ where $\omega
=\sqrt{k_\text{c}/m}$ is the contact-force time scale.

\begin{table}
  \centering
  \begin{tabular}{c|c|c}
    Simulation & Parameter & Value \\
    \hline
    \hline
    \multirow{4}{*}{Fig. 3, $\mathcal{C}=1$} & $N$ & 250 \\
    & $H$ & $2\times10^7$ \\
    & $V_1$ & 0.3333 \\
    & $V_2$ & 0.2333 \\
    \hline
    \multirow{4}{*}{Fig. 3, $\mathcal{C}=-1$} & $N$ & 250 \\
    & $H$ & $4\times10^7$ \\
    & $V_1$ & 2 \\
    & $V_2$ & 0.8 \\
    \hline
    \multirow{4}{*}{Fig. 4, $\mathcal{C}=1$} & $N$ & 130 \\
    & $H$ & $5\times10^7$ \\
    & $V_1$ & 0.3333 \\
    & $V_2$ & 0.2333 \\
    \hline
    \multirow{4}{*}{Fig. 4, $\mathcal{C}=-1$} & $N$ & 160 \\
    & $H$ & $8\times10^7$ \\
    & $V_1$ & 2 \\
    & $V_2$ & 0.8 \\
    \hline
    \multirow{4}{*}{Fig.~\ref{fig:mott-SI}} & $N$ & 200 \\
    & $H$ & $2\times10^8$ ($2\times10^7$) for panel b (c) \\
    & $V_1$ & 0.0667 \\
    & $V_2$ & 0 \\
    \hline
    
  \end{tabular}
  \caption{{Simulation parameters}}
  \label{tab:simparam}
\end{table}

Simulations are initialized with straight, tension-free chains arranged parallel
to the $y$ axis and with horizontal spacing $a$ along $x$. Each simulation is
run for $H$ time steps where $H$ is of the order of $10^7$. To aid
equilibration, simulations begin with an annealing phase in which the
temperature is set to 2--3 times the desired temperature and reduced to the final
temperature via a linear ramp in time for the first $H/2$ steps.
Equilibrium quantities such as density profiles are measured from snapshots of
the bead positions taken at every 1000 to 2000 time steps for the last
quarter of the simulation.

Disorder is implemented by adding to the substrate potential a superposition of
$n_\text{d}$ products of sine functions with random amplitudes, wavelengths, and phases:
\begin{equation}
  \label{eq:random}
  V_\text{d} = \sum_{i=1}^{n_\text{d}} \alpha_i
  \sin\left( \frac{2\pi r_i}{L_x} x + \phi_i\right) \sin\left(\frac{2\pi
      s_i}{L_y}y + \varphi_i\right),
\end{equation}
where amplitude $\alpha_i$ is drawn from a normal distribution with zero mean and standard
deviation $2 \sigma_\text{d} /\sqrt{n_\text{d}}$, and random phases $\phi_i$
and $\varphi_i$ are uniformly distributed in the interval $[0,2\pi)$. The
amplitudes are chosen so that the root-mean-square deviation of the potential
$V_\text{d}$  over the entire substrate matches the desired disorder
strength $\sigma_\text{d}$.

To satisfy the periodic boundary conditions, the random wavelengths are integral
fractions of the simulation box sizes $(L_x,L_y)$, with $r_i$ and $s_i$ drawn uniformly
from integers in the intervals $5 \leq r_i \leq 20$ and $8 \leq s_i \leq 32$
respectively. In Fig.~3, we use $n_\text{d}=8$ and $16$. 

Table~\ref{tab:simparam} shows values of the situation-specific simulation
parameters that have not been defined above.

\section{Computing the scattering intensity profile}

From the numerically averaged distribution of monomer density, we compute the
scattering instensity profile. This quantity is proportional to the static
structure factor $S(\kv)$ for the wavevector $\kv$. The characteristic features
of the scattering profile are the nearest-neighbor  density peaks
situated along the direction separating the chains (i.e., the direction
perpendicular to the average chain orientation). These peaks are signatures of
orientiational order in the chain liquid, and can be compared to the peaks
observed in X-ray scattering profiles for nematic or smectic liquid crystals.
Numerically, we compute the scattering intensity by taking the Fourier transform
of the density-density correlation function, i.e., we compute $\langle \rho(\kv)
\rho(-\kv) \rangle$. We plot the results in Figs.~3b,f of the main text.

\section{Mapping directed line probability distributions to quantum-mechanical wavefunctions} \label{app:comp}

The probability distribution $\Psi(x_1,y|x_0,0)$ describing the statistical
weight of a chain with ends fixed at $x(0)=x_0$ and $x(y)=x_1$ can be
expressed in terms of path integrals as 
\begin{equation}
  \label{eq:partitionfn}
  \Psi(x_1,y|x_0,0) = \int_{x(0)=x_0}^{x(y)=x_1} \mathcal{D}\{x(y)\}
  \exp(-\beta E[x(y)]),
\end{equation}
where $\beta = 1/(k_B T)$ is the inverse thermal energy scale,
and the line energy $E$ is a functional of $x(y)$ of the form
\begin{equation}
  \label{eq:lineenergy}
  E = \int dy \left[\frac{\tau}{2}\left(\frac{dx(y)}{dy}\right)^2 +
    V[x(y)]\right].
\end{equation}
Upon discretizing the path integral above, the Schr\"odinger-like equation
governing the change in partition weights due to the addition of a small element
of length is given by Eq.~(2) in the main text.

The many-body version of main text Eq.~(2) is obtained by replacing the Laplacian operator
$\partial_x^2$ with $\sum_i \partial_{x_i}^2$ and the potential $V(x)$ with a sum over
pairs $\sum_{i<j} V(x_i-x_j)$.

\section{Ground-state dominance and the directed-line Mott insulator}

To further develop the line-quantum mapping, we formally express the solution
to partition function evolution, Eq.~[2],
in terms of the
eigenstates $\psi_n$ and corresponding eigenvalues $\varepsilon_n$ of
$H$~\cite{Nelson2002}:
\begin{equation}
  \label{eq:eigen}
  \Psi(X_0,y|x_0,0) =  \bra{X_0} \left(\sum_n \ket{\psi_n} e^{-\beta \varepsilon_n
    y} \bra{\psi_n}\right) \ket{ x_0}.
\end{equation}
Here, the initial and final ``wavefunctions'' $\ket{X_0}$ and $\ket{x_0}$ are
identified with the probability distributions of the chain position at the
initial and final locations: $\ket{X_0} = \delta(x-X_0)$ in the position basis.
Equation~(\ref{eq:eigen}) highlights a crucial feature of the imaginary-time
evolution: at long distances from the origin, all contributions to $\Psi$ become
exponentially small compared to the ground-state contribution. This situation,
known as \emph{ground state dominance}, simplifies the description of long
directed lines at equilibrium, which is completely captured by the ground
state.

\begin{figure}[t!]
\centering
  \includegraphics{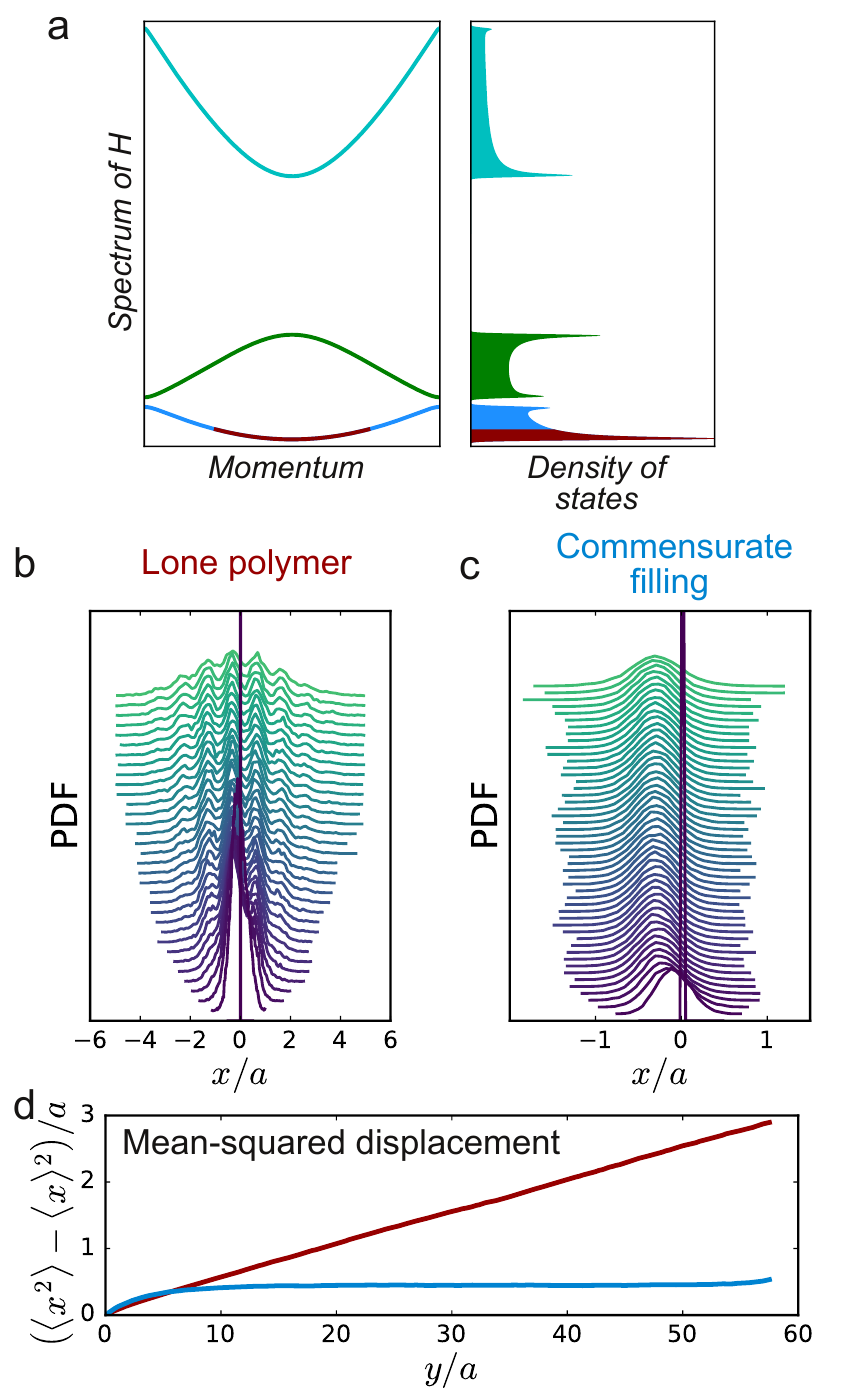}
  \caption{{\bfseries Mott insulator of noncrossing
      directed lines on a periodic substrate.} {\bfseries a,} Example of the
    single-particle band structure for a potential with $V_2=0$ which is
    constant along $y$, and exhibits gaps in the spectrum. The probability
    distribution of a solitary chain is governed by states near the
    zero-momentum eigenstate of the lowest band (red area in {\bfseries a}), and
    wanders over many lattice sites as shown by the spread of the probability
    density function of the chain position with distance from the pinning point
    ({\bfseries b}). By contrast, under commensurate filling (having as many
    chains as potential periods along $x$), the ground state of the
    analogous fermionic system is a completely filled lowest band (blue and red
    areas in {\bfseries a} combined), which puts the system in a Mott-insulating
    state with lines localized to individual potential valleys ({\bfseries
      c}). {\bfseries d,} The localization due to interactions in the
    many-line system (blue curve) tames the random-walk-like evolution of the
    position of a single chain (red curve) as evidenced by the mean square
    displacement of monomer position as a function of distance $y$ (which
    equates to time in the diffusion problem).}
  \label{fig:mott-SI}
\end{figure}

As an example of ground-state dominance for a single chain, we consider the
evolution in the probability density 
function of a single directed line, pinned at $(x_0,0)$, in a periodic potential of
the form $V(x) = V_1 \cos (2\pi x/a)$. The eigenstates of $H$ are Bloch waves of
the form $\psi_{nk}(x) = u_{nk}(x) e^{ikx}$, with cell-periodic functions
$u_{nk}(x)=u_{nk}(x+a)$ of the same periodicity as the potential, and the
normalization $\int dx\,\psi^*_{nk}\psi_{n^\prime k^\prime} =
(2\pi/a)\delta_{n,n^\prime} \delta(k-k^\prime)$ for wavevectors belonging to the
first Brillouin zone (BZ) $k, k^\prime \in [-\pi/a,\pi/a)$. All real
combinations of Bloch waves contribute to the probability evolution; the pinning
condition corresponds to a delta-function probability distribution
$\delta(x-x_0)$ represented as $\braket{x|x_0}$, which has a nonzero overlap
with eigenfunctions across the Brillouin zone. As $y$ increases, however, the
partition sum is dominated by states close to the ground state (red region in
Fig.~\ref{fig:mott-SI}a, which is the $k=0$ eigenstate of the lowest band and
therefore carries equal weight on all lattice sites. As a result, the chain
density becomes more delocalized at larger $y$, as verified in simulations of a single
chain experiencing a periodic potential (Fig.~\ref{fig:mott-SI}b).

As described in the main text, noncrossing interactions among lines can be
exploited to prepare the directed-line system in the analogue of a Mott-insulating
state at a density of one chain per lattice constant $a$. In the absence
of any variation along the (imaginary) time axis, a signature of the
Mott-insulating state is that lines are confined to their respective
potential wells due to repulsion with their neighbours. In Fig.~\ref{fig:mott-SI}c--d,
we verify this fact for the $y$-independent potential. Simulations of a system of
noncrossing lines experiencing the same substrate potential as the single
chain in Fig.~\ref{fig:mott-SI}b, at the density needed to ``fill'' the lowest
band, show that the noncrossing condition confines each line to its
respective well (Fig.~\ref{fig:mott-SI}c), even though the depth of the wells was
not sufficient to confine a single line (Fig.~\ref{fig:mott-SI}b). The diffusive
evolution of the probability distribution with length for the lone chain is
overcome by line-line and line-substrate interactions under
commensurate filling (Fig.~\ref{fig:mott-SI}d).

\section{Many-body position operator for the directed-line system}

The concept of adiabatic pumping is closely connected to the modern theory of
electronic polarization~\cite{King-Smith1993,Resta1994}, which builds on the
realization that changes in polarization and associated current flows in an
electronic system undergoing an adiabatic change are not captured by the
expectation value of the ordinary position operator $x$ which is ill-defined
under periodic boundary conditions. However, they are captured by an
appropriately-defined many-body position operator~\cite{Resta1998,Resta2000}, 
\begin{equation}
  \braket{X} = \frac{L}{2\pi} \text{Im} \ln \bra{\Psi_0} e^{i\frac{2\pi}{L}\sum_{i=1}^N x_i} \ket{\Psi_0},
  \label{eq:pos}
\end{equation}
where $L$ is the system size. 

Below, we apply this operator to the directed-line system. First, let us provide a quantitative description of the Mott-insulating ground state using Wannier functions associated with the lowest band of the periodic potential.
The Wannier functions, $W_{n,R}(x) = W_n(x-R)$ with $W_n(x)=\frac{a}{2\pi}\int_\text{BZ}
dk\,\psi_{nk}(x)e^{-ikx}$, are a set of orthonormal basis
functions for band $n$, each of which is associated with a unit cell at lattice
vector $R=ja$ and exponentially localized around it. For fermions, Wannier
functions are not uniquely defined because of a gauge freedom in defining the
cell-periodic Bloch eigenstates $\ket{u_{n,k}(y)}$, but a real set of Wannier
functions can always be found for isolated bands in one-dimensional periodic
potentials~\cite{Nacbar2007}. When the number of chains equals the number of
unit cells $N$, the fermionic ground state is obtained by assigning one line
to each of the real Wannier functions spanning the lowest band $W_0(x-ja),\, 0
\leq j < N$, and computing a Slater determinant. The many-body ground state
probability distribution of the directed-line system is then the absolute value of the
fermionic ground state~\cite{Girardeau1960}: $\Psi_0^\text{P} = |\Psi_0^\text{F}|,$
where $\Psi_0^\text{P}$ and $\Psi_0^\text{F}$ denote the directed-line and fermionic
ground-state wavefunctions respectively. As a result, the mapping from fermions
to bosons preserves expectation values of operators that commute with the
position wavefunctions.

Continuous changes in the substrate potential shift the centres of mass of the
Wannier functions. A striking result underlying the modern theory of electronic
polarization is that these shifts can be expressed in terms of Berry phases
of the cell-periodic Bloch eigenstates of the corresponding
bands~\cite{Zak1989,King-Smith1993,Resta1994}.
As a result, when the underlying lattice Hamiltonian $H(x+a) = H(a)$
undergoes a periodic change along the $y$ axis $H(x,y) = H(x,y+\lambda)$, the
shift in centre of mass $\langle x \rangle_n = \int dx\, x W_{n,0}^2(x)$ of the
$R=0$ Wannier function over one period is a topological invariant built upon the
Berry phase, quantized by the Chern number $\mathcal{C}_n$.
Note that the Chern number describes a shift in the centre of mass of the
\emph{real} Wannier functions which can be used to build the many-body states of
the directed-line system. 

Now we show that the position operator in Eq.~(\ref{eq:pos}) describes the flow of probability
density of the  chain positions under the imaginary-time Schr\"odinger
evolution, Eq.[2].
For comparison to the quantum case, we use
$t$ in place of $y$ as the coordinate along the tension direction. We work
with eigenstates $\psi_n(x,t)$ of the ``Hamiltonian'' $\beta H = -[(1/2\tau
\beta)\nabla^2-\beta V]$. For convenience we set $\beta =1$ in the remainder of
this section. The Hamiltonian is real, and hence has a set of real orthonormal
eigenstates at each time (instantaneous eigenstates) with corresponding
eigenvalues $\e_n$:
\begin{equation}
  H(t) \psi_n(t)=\e_n(t)\psi_n(t), 
\end{equation}
 and $\langle \psi_m(t) | \psi_n(t) \rangle = \int dx \psi_m(x,t) \psi_n(x,t) =
\delta_{mn}$. We drop the spatial coordinate $x$ in what follows.
If we begin with the decomposition $\Psi(t_0) = \sum_n a_n(t_0) \psi_n(t_0)$,
then we can always find the eigenfunctions at all times $\psi_n(t)$, and the
solution $\psi(t)$ can always be written as as a superposition of these
eigenfunctions, but the key
to describing the evolution lies in finding the quantities $a_n(t)$.  

We write the state at time $t$ as
\begin{equation} \label{eq:wavefunc}
  \Psi(t) = \sum_n a_n(t) \psi_n(t)e^{-\int_{t_0}^t \e_n(t^\prime) dt^\prime},
\end{equation}
so that
\begin{widetext}
\begin{align}
  \dd_t \Psi(t)  &=  -H \Psi(t) \\
  \Rightarrow \sum_n \left(-\e_n a_n \psi_n +\dot{a}_n \psi_n +
  a_n\dot{\psi}_n \right) e^{-\int_{t_0}^t \e_n(t^\prime) dt^\prime} &= -\sum_n \e_n a_n \psi_n
e^{-\int_{t_0}^t \e_n(t^\prime) dt^\prime} \\
  \Rightarrow \sum_n \dot{a}_n \psi_n e^{-\int_{t_0}^t \e_n(t^\prime) dt^\prime}
  &= -\sum_n a_n \dot{\psi_n} e^{-\int_{t_0}^t \e_n(t^\prime) dt^\prime}.
\end{align}
\end{widetext}
Using the orthogonality property, taking a dot product of the equation with
$\psi_m$ gives
\begin{equation}
  \dot{a}_m = -a_m \dotprod{\psi_m}{\dot{\psi}_m} - \sum_{n\neq m} a_n
  \dotprod{\psi_m}{\dot{\psi}_n} e^{-\int_{t_0}^t (\e_n(t^\prime)-\e_m(t^\prime)) dt^\prime}.
\end{equation}
However, since we restrict ourselves to real orthogonal eigenstates, 
$\dd_t\dotprod{\psi_m}{\psi_m} = 0
=\dotprod{\dot{\psi}_m}{\psi_m}+\dotprod{\psi_m}{\dot{\psi}_m} =
2\dotprod{\psi_m}{\dot{\psi}_m} \Rightarrow \dotprod{\psi_m}{\dot{\psi}_m} = 0$.
(In quantum perturbation theory, this condition would be imposed by a ``parallel
  transport'' gauge choice $\langle \psi_m | \dot{\psi}_m \rangle = 0$).
Hence,
\begin{equation}
  \dot{a}_m = - \sum_{n\neq m} a_n
  \dotprod{\psi_m}{\dot{\psi}_n} e^{-\int_{t_0}^t (\e_n(t^\prime)-\e_m(t^\prime)) dt^\prime}.
\end{equation}

If we begin in the ground state: $a_0(t_0) = 1$, then to lowest order we have
$\dot{a}_0 = 0 \Rightarrow a_0(t) = 1$. However the coefficients of the excited
states $m \neq 0$ have a lowest order contribution that is nonzero:
\begin{equation}
  \dot{a}_m = -\dotprod{\psi_m}{\dot{\psi}_0} e^{-\int_{t_0}^t (\e_0(t^\prime)-\e_m(t^\prime)) dt^\prime}.
\end{equation}
with initial condition $a_m(t_0) = 0$. In the adiabatic limit where variations
happen over a time scale $T \to \infty$, we are satisfied with a solution to the
above equations for $a_m(t)$ to lowest order in $1/T$. We obtain this by
integrating once by parts to get
\begin{equation}
  a_m(t) \approx \frac{\dotprod{\psi_m}{\dot{\psi}_0}}{\e_0(t)-\e_m(t)}e^{-\int_{t_0}^t (\e_0(t^\prime)-\e_m(t^\prime)) dt^\prime}.
\end{equation}
The corrections to the above equations go as $\dd_t
\dotprod{\psi_m}{\dot{\psi}_0} \sim \dotprod{\psi_m}{\dot{\psi}_0}/T$, and
higher powers of $1/T$, as required for accuracy in the adiabatic limit. Finally, 
the time-evolved wavefunction at all times can be written using
Eq.~\ref{eq:wavefunc} as
\begin{equation}
  \Psi(t) = e^{-\int_{t_0}^t \e_0(t^\prime) dt^\prime}\left(\psi_0(t) + \sum_{m
      \neq 0} \frac{\dotprod{\psi_m}{\dot{\psi}_0}}{\e_0(t)-\e_m(t)} \psi_m(t)\right).
\end{equation}
At first glance, the time evolution of the ground state appears to depend on the
lowest instantaneous eigenvalue. This is, however, an artifact: although the
imaginary-time Schr\"odinger equation does not conserve probability, any
computations involving the partition weight $\Psi(x,t)$ requires normalization by
the partition sum $Z \equiv \int dx\, \Psi(x,t)$. Therefore, the overall magnitude
of the time-evolved wavefunction is arbitrary and we may choose  a factor that
explicitly conserves net probability integrated over the span of the system:
\begin{equation} \label{eq:tdptwvfn}
  \Psi(t) = \psi_0(t) + \sum_{m
      \neq 0} \frac{\dotprod{\psi_m}{\dot{\psi}_0}}{\e_0(t)-\e_m(t)} \psi_m(t),
\end{equation}
for which $\dotprod{\Psi}{\Psi} = 1 + O(1/T^2)$.

We are interested in the probability distribution of points in the interior of a
directed line at $y = t$, far away from the ends at $y = 0$ and $y = L$, which is
described purely by the ground state of the Hamiltonian regardless of pinning
conditions. In addition to the partial partition weight $\Psi(x,t|x_0,0)$ of
finding the line at position $x$ at $y$-coordinate $t$, we need to consider
the weight $\Gd(x,t|x_L,L)$ associated with conformations connecting the
interior point at $t$ to the pinning point at $L$. The complementary partition
function is governed by the equation $-\partial_t \Gd = [(1/2\tau
\beta)\nabla^2-\beta V] \Gd = -\beta H \Gd$. 
The dagger does not signify conjugation; $\Psi$ and $\Gd$ are different functions.
However the notation is suggestive and the situation parallels that of a
wavefunction and its conjugate in the quantum mechanical picture. Repeating the
time-dependent perturbation theory calculation provides the following expression
for $\Gd(t)$:
\begin{equation} \label{eq:tdptwvfn2}
  \Gd(t) = \psi_0(t) - \sum_{m
      \neq 0} \frac{\dotprod{\psi_m}{\dot{\psi}_0}}{\e_0(t)-\e_m(t)} \psi_m(t),
\end{equation}

The spatial density distribution at $y$-coordinate $t$ is written as:
\begin{equation}
  \label{eq:prob}
  \rho(x,t) = \frac{1}{Z} \Psi(x,t|x_0,0)\Gd(x,t|x_L,L),
\end{equation}
where $Z = \int dx\, \Psi(x,r_0;t) \Gd(x,r_1;t) = \Psi(x_L,L|x_0,0)$ is the full
partition function of the chain and therefore independent of $t$. We can treat
$Z$ as a time-independent normalization, and we will not write it
explicitly in what follows. The time evolution of the
spatial density produces a ``probability current density'' $j$ through the continuity
equation
\begin{align}
  \dd_t \rho(x,t) + \dd_x j(x,t) &= 0 \\
  \Rightarrow \dd_x j &= - \Psi \dd_t \Gd - \Gd \dd_t \Psi \\
  = \frac{1}{2\tau\beta} \left(\Psi \dd_x^2 \Gd - \Gd \dd_x^2 \Psi\right) &-\beta
  V\left(\Psi \Gd - \Gd \Psi\right) \\
  \Rightarrow j &= \frac{1}{2\tau\beta}\left( \Psi \dd_x \Gd - \Gd \dd_x \Psi\right).
\end{align}
(we reintroduce the general $\beta$ from now on).

The shift in expected position of the chain upon traversing a period $T$
in ``time'' (i.e. distance along the chain) is now obtained by integrating the
current density over a period and over all space.
Substituting the perturbative expressions for $\Psi$ and $\Gd$ into the current
equation, we get
\begin{equation}
  j(x,t) = \frac{1}{2\tau\beta}\sum_{m \neq 0}
  \frac{\dotprod{\psi_m}{\dot{\psi}_0}}{\e_0(t)-\e_m(t)}
  \left[ 2 \psi_m \dd_x \psi_0 - 2 \psi_0 \dd_x \psi_m \right]. 
\end{equation}
Integrating over space to get the average current, we have
\begin{equation} \label{eq:intcurrent}
  J(t) = \frac{1}{L}\int dx\, j(x,t) = \frac{2}{\tau\beta} \sum_{m\neq 0}
  \frac{\dotprod{\dd_x\psi_0}{\psi_m} \dotprod{\psi_m}{\dd_t \psi_0}}{\e_0-\e_m},
\end{equation}
where we have used the orthogonality $\dotprod{\psi_m}{\psi_0} = 0$ which implies
$\dotprod{\psi_m}{\dd_x \psi_0} +\dotprod{\dd_x \psi_m}{\psi_0} = 0$.

Using the fact that the instantaneous eigenstates $\psi_i$ can always be chosen to be real for
real Hamiltonians, the right hand side of Eq.~\ref{eq:intcurrent} may be written
as
\begin{equation} \label{eq:intcurrent2}
  J(t) = \frac{1}{\tau\beta} \sum_{m\neq 0}
  \left(\frac{\dotprod{\dd_x\psi_0}{\psi_m} \dotprod{\psi_m}{\dd_t
        \psi_0}}{\e_0-\e_m} + \right.
      \left. \frac{\dotprod{\dd_t\psi_0}{\psi_m}
      \dotprod{\psi_m}{\dd_x \psi_0}}{\e_0-\e_m}\right),
\end{equation}
Upon identifying the quantum-mechanical momentum operator $p$ with
$-i\hbar \partial_x$ and substituting $\hbar^2/2m$ for $1/2\tau\beta$, the right
hand side of Eq.~\ref{eq:intcurrent2} is identical to the average flow of
current under adiabatic evolution of a quantum electronic system with the same
potential $V(x)$~\cite{Thouless1983,Niu1984,Resta1998}, which forms the basis
for the theories of polarization and quantized charge pumping. Furthermore, for
adiabatic evolution of the Hamiltonian, the current is the time-derivative of
the expectation value of the many-body position operator:
\begin{equation}
  \label{eq:dtpos}
  J(t) = \frac{1}{L}\frac{d}{dt}\langle X \rangle.
\end{equation}
Like the position operator, the probability flow therefore depends only on the
square modulus of the instantaneous ground-state wavefunction~\cite{Resta1998}.
Since the instantaneous adiabatic ground states of the directed line system are
simply the absolute values of the corresponding ground states of the electron
system and share the same form of their probability evolution over time, the
equivalence of the two problems has been established.

The equivalence of the probability current with the quantum-mechanical system
and the existence of exponentially localized Wannier functions for the directed line
system also provides a route to rigorously proving the robustness of the tilt
under many-body interactions and substrate disorder, following the techniques of
Niu and Thouless~\cite{Niu1984}.

\section{Descriptions of SI Videos}

\subsection{SI Video 1}
One chain fluctuating in a two-dimensional external potential. 
This chain does not exhibit a topologically protected tilt because
there is no band gap to (diffusive) excitations. 
Instead, the chain diffuses freely over many lattice sites.

\subsection{SI Video 2}
Nine chains fluctuating in a two-dimensional external potential
with an incommensurate system size (corresponding to ten periods of the potential
in the horizontal direction).
In this video, the single vacancy (i.e., unfilled lattice site) diffuses freely
much like the single chain in SI Video 1.
Despite strong interactions, there is no gap to excitations
in this case of incommensurate lattice filling.
Again, the chains do not exhibit a topologically protected tilt. 

\subsection{SI Video 3}
Ten chains fluctuating in a two-dimensional external potential
with a \emph{commensurate} system size, having
exactly ten periods of the potential.
Due to a gap in excitations, the individual chains
no longer freely diffuse but are instead confined
to their potential wells by strong repulsive interactions with their neighbors (c.f., SI Figure 1).
In this case of commensurate filling the system does exhibit a topologically protected tilt.


%

\end{document}